\documentclass[twoside]{article}
\usepackage{fleqn,espc}
\usepackage{graphicx}

\title{Renormalizability and searching for $Z'$ gauge bosons}

\author{
 A.V. Gulov
 \address{Dniepropetrovsk National University \\
          Dniepropetrovsk 49050, Ukraine}
 \thanks{Email gulov@ff.dsu.dp.ua}
 and V.V. Skalozub
 \addressmark
 \thanks{Email skalozub@ff.dsu.dp.ua}
 }

\begin{document}

\begin{abstract}
New observables for searching for $Z'$ virtual states in processes
$e^+e^-\to\bar{f}f$ are proposed. They are based on the relations
following due to renormalizability of an underlying theory between
the parameters of the low-energy effective Lagrangian. Two types
of the $Z'$ interactions are considered. The observables allow to
pick up uniquely the signals of both $Z'$ bosons. The search for
the $Z'$ at future linear colliders and the treating of the LEP
data are discussed.
 \vspace{1pc}
\end{abstract}

\maketitle

\section{INTRODUCTION}

Among various objectives of the recently finished LEP experiments
an important place was devoted to searching for signals of new
physics beyond the energy scale of the standard model (SM).
Reports on these results are adduced partly in the literature
\cite{osaka}. In the present note we are going to discuss the
problem of searching for the heavy $Z'$ gauge boson. This particle
is a necessary element of the different models extending the SM.
Low limits on its mass following from the analysis of variety of
popular models are found to be in the energy intervals 500--2000
GeV \cite{osaka} (see Table 1 which reproduces Table 9 of Ref.
\cite{osaka}). As it is seen, the values of $m_{Z'}$ (as well as
the parameters of interactions with the SM particles) are strongly
model dependent. Therefore, it seems reasonable to find some model
independent signals of this particle. To elaborate that general
principles of field theory must be taken into consideration giving
a possibility to relate the parameters of different scattering
processes. Then, one is able to introduce variables, convenient
for the model independent search for $Z'$ (or other heavy states).
These ideas were used in \cite{ZprTHDM} in order to introduce the
model-independent sign definite variables for $Z'$ detection in
scattering processes with $\sqrt{s}\simeq 500$ GeV.

As it has been pointed out in \cite{ZprTHDM}, some parameters of
new heavy fields can be related by using the requirement of
renormalizability of the underlying model remaining in other
respects unspecified. The relations between the parameters of new
physics due to the renormalizability were called the
renormalization group (RG) relations. In Ref. \cite{ZprTHDM} the
RG relations for the low-energy couplings of the SM fields to the
heavy neutral gauge boson ($Z'$ boson) have been derived. They
predict two possible types of $Z'$ particles, namely, the chiral
and the Abelian $Z'$ ones. Each type is described by a few
couplings to the SM fields. Therefore, taking into account the RG
correlations between the $Z'$ couplings, one is able to introduce
observables which uniquely pick up the $Z'$ virtual state
\cite{ZprTHDM}. In the present paper we discuss these observables
and the possibility of searching for $Z'$ at the present day and
future $e^+e^-$ colliders.

\begin{table}
  \centering
  \caption{95\% confidence level lower limits on the $Z'$
  mass for $\chi$, $\psi$, $\eta$, L--R models \cite{models} and
  for the Sequential Standard Model (SSM) \cite{SSM}.}
\begin{tabular}{|c|c|c|c|c|c|}
  \hline
  Model & $\chi$ & $\psi$ & $\eta$ & L--R & SSM \\
  \hline \hline
  $m^{\rm limit}_{Z'}, {\rm GeV}/c^2$ &
   630 & 510 & 400 & 950 & 2260 \\
  \hline
\end{tabular}\end{table}

\section{$Z'$ COUPLINGS TO FERMIONS}

As it was pointed out in Ref. \cite{ZprTHDM}, to derive RG
correlations between the parameters of the $Z'$ interactions with
light particles one must specify the model describing physics at
low energies. For instance, the minimal SM with the one scalar
doublet can be chosen. However, due to the lack of information
about scalar fields, models with an extended set of light scalar
particles can be also considered. Below we choose the
two-Higgs-doublet model (THDM) \cite{THDM} as the low-energy
theory (notice, the minimal SM is the particular case of the
THDM).

To derive the RG relations one has to introduce the
parametrization of $Z'$ couplings to the SM fields. Since we are
going to account of the $Z'$ effects in the low-energy
$e^+e^-\to\bar{f}f$ processes in lower order in $m^{-2}_{Z'}$, the
linear in $Z'$ interactions with the SM fields are of interest,
only. The renormalizability of the underlying theory and the
decoupling theorem \cite{decoupling} guarantee the dominance of
renormalizable $Z'$ interactions at low energies. The interactions
of the non-renormalizable types, generated at high energies due to
radiation corrections, are suppressed by the inverse heavy mass.
The SM gauge group $SU(2)_L\times U(1)_Y$ is considered as a
subgroup of the underlying theory group. So, the mixing
interactions of the types $Z'W^+W^-$, $Z'ZZ$, ... are absent at
the tree level. Such conditions are usually used in the literature
\cite{EL1} and lead to the following parametrization of the linear
in $Z'$ low-energy vertices:
\begin{eqnarray}\label{1}
 {\cal L}&=&
 \sum\limits_{i=1}^2
  \left|\left(
  D^{{\rm ew,} \phi}_\mu -
  \frac{i\tilde{g}}{2}\tilde{Y}(\phi_i)\tilde{B}_\mu
  \right)\phi_i\right|^2
 +
 \nonumber\\&&
 i\sum\limits_{f=f_L,f_R}\bar{f}{\gamma^\mu}
  \left(
  D^{{\rm ew,} f}_\mu -
  \frac{i\tilde{g}}{2}\tilde{Y}(f)\tilde{B}_\mu
  \right)f,
\end{eqnarray}
where $\phi_i$ $(i=1,2)$ are the scalar doublets, $\tilde{g}$ is
the charge corresponding to the $Z'$ gauge group, $D^{{\rm
ew,}\phi}_\mu$ and $D^{{\rm ew,}f}_\mu$ are the electroweak
covariant derivatives, $\tilde{B}_\mu$ denotes the massive $Z'$
field before the spontaneous breaking of the electroweak symmetry,
and the summation over the all SM left-handed fermion doublets,
$f_L =\{(f_u)_L, (f_d)_L\}$, and the right-handed singlets, $f_R =
(f_u)_R, (f_d)_R$, is understood. Diagonal $2\times 2$ matrices
$\tilde{Y}(\phi_i)$, $\tilde{Y}(f_L)$ and numbers $\tilde{Y}(f_R)$
are unknown $Z'$ generators characterizing the model beyond the
SM.

The one-loop RG relations for the above introduced $Z'$ vertices
(\ref{1}) have been obtained in Ref. \cite{ZprTHDM}. As it was
shown, two different types of the $Z'$ generators are compatible
with the renormalizability of the underlying theory. The first
type, called the chiral $Z'$, describes the $Z'$ boson which
couples to the SM doublets, only. The corresponding generators
have the zero traces:
\begin{eqnarray}\label{chiral}
 &&\tilde{Y}(\phi_i)=
 -\tilde{Y}_{\phi}
 \left(\begin{array}{cc}
 1 & 0 \\ 0 & -1
 \end{array}\right),
 \nonumber\\&&
 \tilde{Y}(f_L)=
 \tilde{Y}_{L,f_u}
 \left(\begin{array}{cc}
 1 & 0 \\ 0 & -1
 \end{array}\right),
 \nonumber\\&&
 \tilde{Y}(f_R)=0.
\end{eqnarray}
The second type is the Abelian $Z'$ boson:
\begin{eqnarray}\label{abelian}
 &&\tilde{Y}(\phi_i)=
 \tilde{Y}_{\phi}
 \left(\begin{array}{cc}
 1 & 0 \\ 0 & 1
 \end{array}\right),
 \nonumber\\&&
 \tilde{Y}(f_L)=
 \tilde{Y}_{L,f}
 \left(\begin{array}{cc}
 1 & 0 \\ 0 & 1
 \end{array}\right),
 \nonumber\\&&
 \tilde{Y}(f_R)=
 \tilde{Y}_{L,f}+2T^3_f \tilde{Y}_{\phi},
\end{eqnarray}
where $T^3_f$ is the third component of the fermion weak isospin.
The relations (\ref{abelian}) ensure, in particular, the
invariance of the Yukawa terms with respect to the effective
low-energy $\tilde{U}(1)$ subgroup corresponding to the $Z'$
boson. As it follows from the relations, the couplings of the
Abelian $Z'$ to the axial-vector fermion currents have the
universal absolute value proportional to the $Z'$ coupling to the
scalar doublets.

The derived relations (\ref{chiral})--(\ref{abelian}) are
model-independent ones. They hold in the THDM as well as in the
minimal SM. As it is seen from relations
(\ref{chiral})--(\ref{abelian}), only one parameter for each SM
doublet remains arbitrary. The rest parameters are expressed
through them. A few number of independent $Z'$ couplings gives the
possibility to introduce the observables convenient for detecting
uniquely the $Z'$ signals in experiments. In what follows, we
consider the searching for the $Z'$ at future linear $e^+e^-$
colliders and treating the obtained at LEP data taking into
account the RG relations (\ref{chiral})--(\ref{abelian}).

\section{$Z'$ SIGNALS AT FUTURE $e^+e^-$ COLLIDERS}

The model-independent $Z'$ searches at $e^+e^-$ colliders are
intensively discussed in the literature (see, for instance, the
reports \cite{leike,riemann}). The LEP experiments at energies
$\sqrt{s}\simeq 200$ GeV were completed recently. The analysis of
the obtained data will constrain, in particular, the possible $Z'$
signals. In this regard, the RG relations
(\ref{chiral})--(\ref{abelian}) can be used to introduce the
observables convenient for the detection of the $Z'$ signals in
the electron-positron annihilation into the fermion pairs
\cite{ZprTHDM,obs}. Before discussing the LEP case, we consider
the observables for future linear colliders with the
center-of-mass energy $\sqrt{s}\ge 500$ GeV. At energies
$\sqrt{s}\ge 500$ GeV the observables can be simplified
significantly, since the $Z$-boson mass can be neglected and all
the $Z'$ effects can be described by the four-fermion contact
interactions induced by the heavy $Z'$ exchange. The magnitude of
these contact interactions is independent of the center-of-mass
energy, therefore any value of $\sqrt{s}$ gives predictions close
to the $m_Z\to 0$.

Consider the process $e^+e^-\to V^\ast\to\bar{f}f$ ($f\not=e,t$)
with the neutral vector boson exchange ($V=A,Z,Z'$). We assume the
non-polarized initial- and final-state fermions. Since the $t$
quark is not considered, the fermions can be treated as massless
particles, $m_f\sim 0$. In this approximation the left-handed and
the right-handed fermions can be substituted by the helicity
states, which we mark as $\lambda$ and $\xi$ for the incoming
electron and the outgoing fermion, respectively
($\lambda,\xi=L,R$).

The differential cross section of the process $e^+e^-\to
V^\ast\to\bar{f}f$ deviates from its SM value by a quantity of
order $m^{-2}_{Z'}$:
\begin{eqnarray}\label{eq7}
 \Delta\frac{d\sigma_f}{d\cos\theta}&=&
 \frac{1}{16\pi s} \mbox{Re}\Big[
 {\cal A}^\ast_{\rm SM}
 \nonumber\\&&\times
 \left({\cal A}_{Z^\prime} +
 \left.\frac{d{\cal A}_Z}{d\theta_0}\right|_{\theta_0 =0}
 \theta_0 \right)\Big],
 \nonumber\\
 {\cal A}_{\rm SM}&=&{\cal A}_A + {\cal A}_Z(\theta_0 =0),
\end{eqnarray}
where $\theta$ denotes the angle between the momentum of the
incoming electron and the momentum of the outgoing fermion, ${\cal
A}_V$ is the Born amplitude of the process, and $\theta_0\sim
m^2_W/m^2_{Z'}$ is the $Z$--$Z'$ mixing angle. The leading
contribution comes from the interference between the $Z'$ exchange
amplitude, ${\cal A}_{Z'}$, and the SM amplitude, ${\cal A}_{\rm
SM}$, whereas the $Z$--$Z'$ mixing terms are suppressed by the
additional small factor $m^2_Z/s$. Notice that the deviation
$\Delta d\sigma_f/d\cos\theta$ depends on the center-of-mass
energy through the quantity $m^2_Z/s$, only.

To take into consideration the correlations (\ref{chiral}),
(\ref{abelian}) let us introduce the observable $\sigma_f(z)$
defined as the difference of cross sections integrated in some
ranges of the scattering angle $\theta$, which will be specified
below \cite{ZprTHDM,obs}:
\begin{eqnarray}\label{eq8}
 \sigma_f(z)
 &\equiv&\int\nolimits_z^1
  \frac{d\sigma_f}{d\cos\theta}d\cos\theta
 -\int\nolimits_{-1}^z
  \frac{d\sigma_f}{d\cos\theta}d\cos\theta
 \nonumber\\&=&
 \sigma^T_f\left[ A^{FB}_f\left(1-z^2\right)
 -\frac{z}{4}\left(3+z^2\right)\right],
\end{eqnarray}
where $z$ stands for the boundary angle, $\sigma^T_f$ denotes the
total cross section and $A^{FB}_f$ is the forward-backward
asymmetry of the process. The idea of introducing the
$z$-dependent observable (\ref{eq8}) is to choose the value of the
kinematic parameter $z$ in such a way that to pick up the
characteristic features of the $Z'$ signals. In the next sections
we will consider the deviations from the SM predictions,
$\Delta\sigma_f(z)$, induced by the chiral and the Abelian $Z'$
bosons.

\subsection{Chiral $Z'$}

In case of the chiral $Z'$ the deviation $\Delta\sigma_f(z)$ can
be expressed as
\begin{eqnarray}\label{obs1}
 \Delta\sigma_f(z)&=&
 \frac{\alpha_{\rm em}\tilde{g}^2}{32m^2_{Z'}}
 \left(
 {\cal G}^f_1(z,s) \tilde{Y}_{L,f}\tilde{Y}_{L,e}
 \right.\nonumber\\&&
 +{\cal G}^f_2(z,s) \tilde{Y}_{L,f}\tilde{Y}_\phi
 \nonumber\\&&\left.
 +{\cal G}^f_3(z,s) \tilde{Y}_{L,e}\tilde{Y}_\phi
 \right),
\end{eqnarray}
where $\alpha_{\rm em}$ is the fine structure constant, and ${\cal
G}^f_i(z,s)$ are the factors dependent on the SM parameters, only.
The leading contribution to the quantity $\Delta\sigma_f(z)$ comes
from the first term in the sum, since the functions ${\cal
G}^f_2(z,s)$ and ${\cal G}^f_3(z,s)$ originate from $Z$--$Z'$
mixing and contain the additional small factor $m^2_Z/s$. So, for
the center-of-mass energies $\sqrt{s}\ge 500$ GeV the deviation
$\Delta\sigma_f(z)$ is computed to within 2\% in the form:
\begin{eqnarray}\label{eq9}
 \Delta\sigma_f(z)&\simeq&
 \frac{\alpha_{\rm em}\tilde{g}^2}{32m^2_{Z'}}
 {\cal G}^f_1(z,s) \tilde{Y}_{L,f}\tilde{Y}_{L,e},
 \nonumber\\
 {\cal G}^f_1(z,s)&\simeq&
 \frac{4 T^3_f N_f}{3}
 \left(1 +|Q_f| +\ldots\right)
 \nonumber\\&&\times
 \left(1 -z -z^2 -\frac{z^3}{3}\right),
\end{eqnarray}
where $N_f$ is the number of colors, $Q_f$ is the fermion charge
in the positron charge units, and dots stand for the terms of
order $O(1-4\sin^2{\theta_W},m^2_Z/s)$ ($\theta_W$ is the Weinberg
angle).

As it is seen, the quantity $\Delta\sigma_f(z)$ is proportional to
the same polynomial in $z$ for any final fermion state. As a
consequence of these factorization, the observable
$\Delta\sigma_f(z)$ is completely determined by the deviation of
the total cross section. This is the distinguished property of the
chiral $Z'$ signal.

Comparing the deviations $\Delta\sigma_f(z)$ for the fermions
which belong to the same left-handed isodoublet, $\{f_u,f_d\}$,
one finds that the ratio $\Delta\sigma_{f_u}(z)/
\Delta\sigma_{f_d}(z)$ is independent of the boundary angle $z$.
It equals to 5/4 for quarks and 1/2 for leptons in lower order in
small parameters $1-4\sin^2{\theta_W}\simeq 0.08$ and $m^2_W/s$.

At $z=2^{2/3}-1\simeq 0.5874$ the quantity $\Delta\sigma_f(z)$
becomes zero. In other words, the chiral $Z'$ signal can be
kinematically suppressed by choosing this value of the boundary
angle. On the other hand, among the deviations $\Delta\sigma_f(z)$
computed at different boundary angles $z$, the deviation of the
total cross section is maximal.

\subsection{Abelian $Z'$}

\begin{figure}[htb]
\includegraphics[bb= 15 0 585 520,width=55mm]{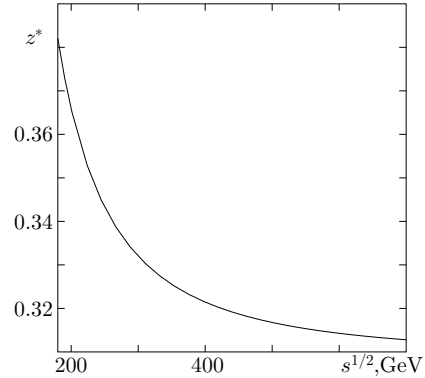}
\caption{$z^\ast$ as the function of the center-of-mass energy.}
\label{fig:zast}
\end{figure}

Now, let us consider the observables for the Abelian $Z'$ signals at
future linear colliders. In this case the quantity (\ref{eq8}) can
be written as follows
\begin{eqnarray}\label{obs2}
 \Delta\sigma_f(z)
 &=& \frac{\alpha_{\rm em}\tilde{g}^2}{16m^2_{Z'}}
 \left(
 {\cal F}^f_0(z,s)a^2_{Z'}
 \right.\nonumber\\&&
 +{\cal F}^f_1(z,s)v^f_{Z'}v^e_{Z'}
 +{\cal F}^f_2(z,s)v^f_{Z'}|a_{Z'}|
 \nonumber\\&&\left.
 +{\cal F}^f_3(z,s)v^e_{Z'}|a_{Z'}|
 \right).
\end{eqnarray}
where $v^f_{Z'}\equiv(\tilde{Y}_{L,f}+\tilde{Y}_{R,f})/2$ and
$a^f_{Z'}\equiv (\tilde{Y}_{R,f}-\tilde{Y}_{L,f})/2=
T^3_f\tilde{Y}_{\phi}$ are the $Z'$ couplings to the vector and
the axial-vector fermion currents. Functions ${\cal F}^f_i(z,s)$
are determined by the SM quantities and depend on the fermion type
throw the charge $Q_f$ and the number of colors $N_f$, only. They
are independent of the fermion generation. The leading
contributions to the lepton factors ${\cal F}^l_2(z,s)={\cal
F}^l_3(z,s)$ equal to zero. So, by choosing the boundary angle
$z^\ast$ to be the solution to the equation ${\cal
F}^l_1(z^\ast,s)=0$, one can switch off three lepton factors
${\cal F}^l_i(z,s)$ $(i=1,2,3)$ simultaneously. Note that the
function $z^\ast(s)$ is the decreasing function of energy. This is
shown in Fig. \ref{fig:zast} for energies $\sqrt{s}=200$--700 GeV.

By choosing $z=z^\ast$, we obtain the sign definite observable
\begin{eqnarray}
 \Delta\sigma_l(z^\ast)&\simeq&
 \frac{\alpha_{\rm em}\tilde{g}^2}{16 m^2_{Z^\prime}}
 {\cal F}^l_0(z^\ast,s)a^2_{Z^\prime}
 \nonumber\\
 &\simeq&
 -0.10\frac{\alpha_{\rm em}\tilde{g}^2\tilde{Y}^2_\phi}
 {16 m^2_{Z^\prime}}<0.
\end{eqnarray}
The quantity $\Delta\sigma_l(z^\ast)$ is negative and the same for
the all types of SM charged leptons. This is the model-independent
signal of the Abelian $Z'$ boson.

\begin{figure}[htb]
\includegraphics[bb= 0 0 600 600,width=55mm]{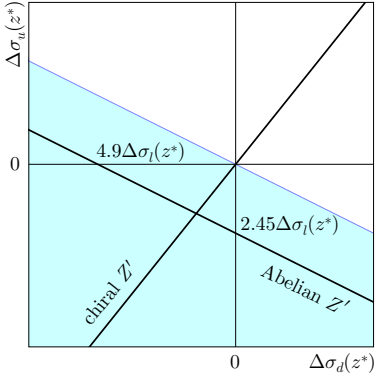}
\caption{Signals of the chiral and the Abelian $Z'$ in the plane
of the observables for quarks of the same generation. The shaded
area represents the Abelian $Z^\prime$ signal for all possible
values of the axial-vector couplings $a^f_{Z^\prime}$.}
\label{fig:quarks}
\end{figure}

For the energy values in the range $\sqrt{s}\ge 500$ GeV one also
is able to introduce the sign definite observables for the quarks of the
same generation:
\begin{eqnarray}
\Delta\sigma_q(z^\ast) &\equiv&\Delta\sigma_{q_u} +0.5
\Delta\sigma_{q_d}
 \nonumber\\
&\simeq& 2.45\Delta\sigma_l(z^\ast)<0.
\end{eqnarray}
Thus, the signal of the Abelian $Z'$ boson can be detected from
the quark variables $\Delta\sigma_{q_u}(z^\ast)$ and
$\Delta\sigma_{q_d}(z^\ast)$ which are dependent. Moreover, they
are related to the lepton observable $\Delta\sigma_l(z^\ast)$.
Possible values of the quark variables in the plane
$\Delta\sigma_{q_u}(z^\ast)$-- $\Delta\sigma_{q_d}(z^\ast)$ are to
be at a straight line crossing the axes in the points
$\Delta\sigma_{q_u}(z^\ast)=2.45\Delta\sigma_l(z^\ast)$ and
$\Delta\sigma_{q_d}(z^\ast)=4.9\Delta\sigma_l(z^\ast)$,
respectively (see Fig. \ref{fig:quarks}). As the chiral $Z'$ boson
is concerned, experimental data have to be at some point on the
different straight line
$\Delta\sigma_{q_u}(z^\ast)=5\Delta\sigma_{q_d}(z^\ast)/4$. This
fact is very important distinguishable feature of the variables
$\Delta\sigma_l(z^\ast)$, $\Delta\sigma_q(z^\ast)$, which select
the $Z'$ boson signals in the processes $e^+e^-\to\bar{f}f$. As a
result, introduced observables are perspective in the
model-independent searching for the $Z'$ signals at future linear
colliders.

\section{$Z'$ SIGNALS AT LEP ENERGIES}

\begin{figure}[htb]
\vspace{9pt}
\includegraphics[bb= 25 0 565 520,width=55mm]{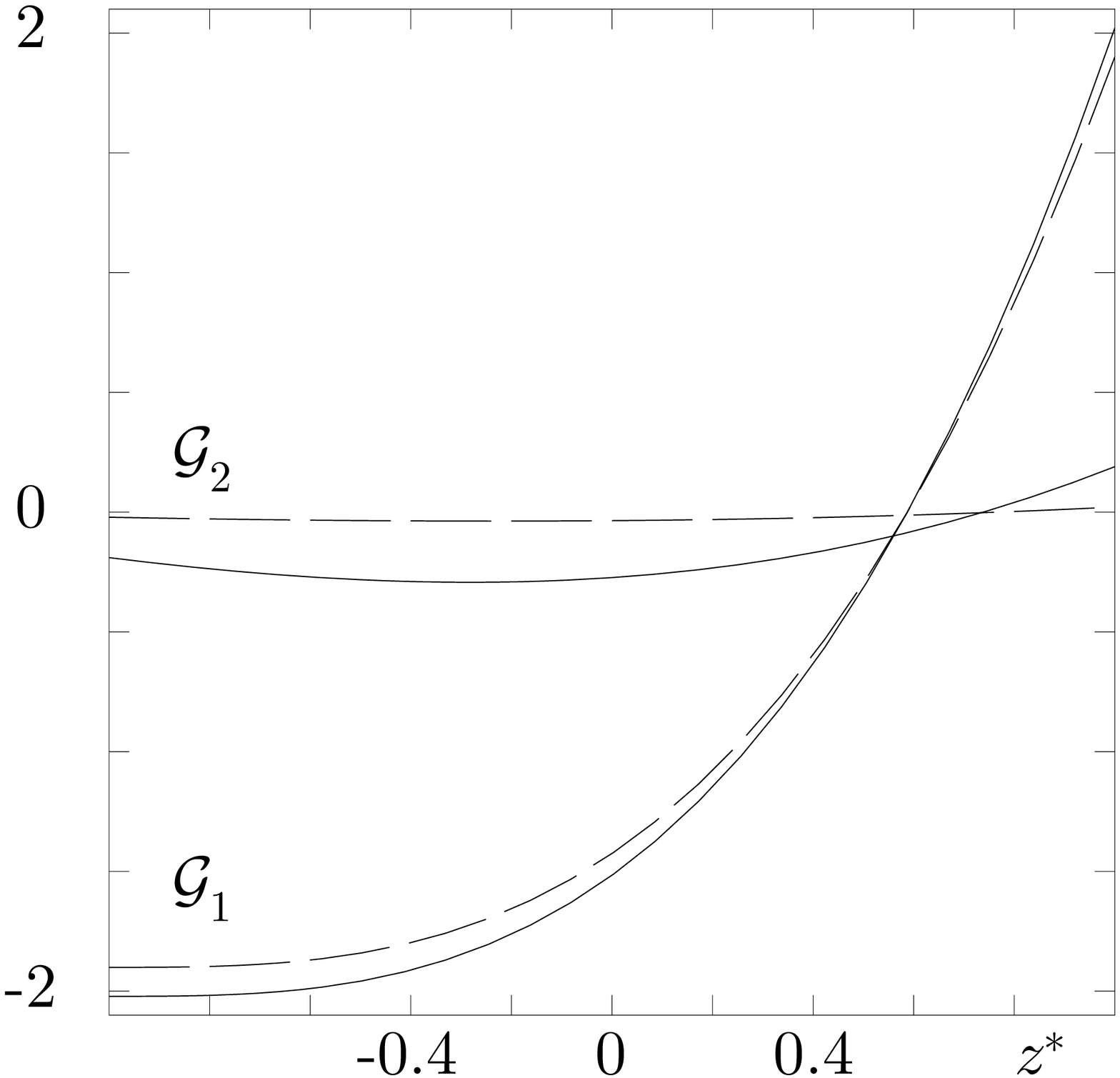}
\caption{The lepton factors
 ${\cal G}^l_1$, ${\cal G}^l_2={\cal G}^l_3$,
 entering the chiral $Z'$ observables $\Delta\sigma_l(z)$,
 computed at the center-of-mass energies
 $\sqrt{s}=200$ GeV (the solid curves) and
 $\sqrt{s}=500$ GeV (the dashed curves).}
 \label{fig:ch}
\end{figure}

\begin{figure}[htb]
\includegraphics[bb= 25 0 565 520,width=55mm]{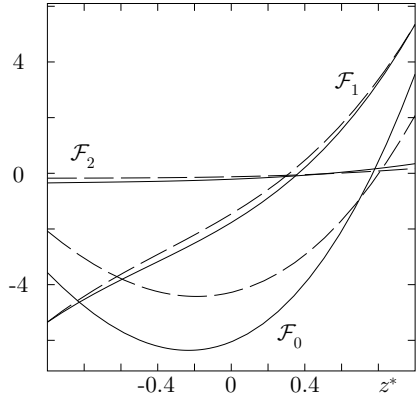}
\caption{The lepton factors
 ${\cal F}^l_0$, ${\cal F}^l_1$, ${\cal F}^l_2={\cal F}^l_3$,
 entering the Abelian $Z'$ observables $\Delta\sigma_l(z)$,
 computed at the center-of-mass energies
 $\sqrt{s}=200$ GeV (the solid curves) and
 $\sqrt{s}=500$ GeV (the dashed curves).}
 \label{fig:ab}
\end{figure}

Now, let us discuss the applicability of the introduced
observables to analyse the LEP data. At the LEP center-of-mass
energies $\sqrt{s}\sim 200$ GeV the effects of the $Z$--$Z'$
mixing become more pronounced and the observables (\ref{obs1}) and
(\ref{obs2}) become more complicated. In what follows, we consider
the lepton observables $\Delta\sigma_l(z)$, only. Some lepton
factors ${\cal G}^l_i(z)$ and ${\cal F}^l_i(z)$, entering the
quantities (\ref{obs1}) and (\ref{obs2}), are dependent: ${\cal
G}^l_2(z,s)={\cal G}^l_3(z,s)$ and ${\cal F}^l_2(z,s)={\cal
F}^l_3(z,s)$. In Figs. \ref{fig:ch}--\ref{fig:ab} we compare the
factors for the chiral $Z'$ and the Abelian $Z'$ computed at the
energies $\sqrt{s}= 200$ GeV and $\sqrt{s}= 500$ GeV.


As it is seen, at the LEP energies the contributions of the chiral
factors ${\cal G}^l_2(z)={\cal G}^l_3(z)$ are of order 10--15\% of
${\cal G}^l_1(z)$. Therefore, they cannot be omitted, as it was resonable
done for energies $\sqrt{s} = 500$ GeV. These contributions spoil the
factorization of the dependence on the boundary angle $z$ for the
observable $\Delta\sigma_l(z)$. Therefore, the proportionality of
$\Delta\sigma_l(z)$ to the total cross-section deviation
$\Delta\sigma^T_l$ does not hold.

In Fig. \ref{fig:ab} we show the factors ${\cal F}^l_i(z,s)$
computed at energies $\sqrt{s}=200$ and 500 GeV. As it is
occurred, the only factor ${\cal F}^l_0(z)$ at $a^2_{Z'}$ changes
significantly. Other functions ${\cal F}^l_1(z,s)$, ${\cal
F}^l_2(z,s)={\cal F}^l_3(z,s)$ contribute less than 2\%. Hence, as
the Abelian $Z'$ is concerned, one is able to introduce the
sign-definite lepton observables $\Delta\sigma_l(z^\ast)$ at the
LEP energies. These observables differ from ones at $\sqrt{s}=500$
GeV by the larger value of the boundary angle $z^\ast=0.366$
($z^\ast=0.317$ at $\sqrt{s}=500$ GeV).

In fact, the quantities $\Delta\sigma_l(z^\ast)$ depend on one
unknown $Z'$ parameter, $\tilde{g}^2\tilde{Y}^2_\phi/m^2_{Z'}$,
only:
\begin{equation}
\Delta\sigma_l(z^\ast)\simeq
 -0.528
 \frac{\alpha_{\rm em}\tilde{g}^2\tilde{Y}^2_\phi}
 {16 m^2_{Z'}}.
\end{equation}
Moreover, this parameter determines the deviation of the
$\rho$-parameter, $\rho\equiv m^2_W/(m^2_Z\cos^2{\theta_W})$, from
unit \cite{ZprTHDM,obs}:
\begin{equation}\label{12}
 \frac{\tilde{g}^2\tilde{Y}^2_\phi}{m^2_{Z'}}
 \simeq
 \frac{4\pi\alpha_{\rm em}}{m^2_W\sin^2{\theta_W}}
 \left( \rho - 1\right).
\end{equation}
As a result, the Abelian $Z'$ signal is described by the
model-independent relations:
\begin{eqnarray}
 \Delta\sigma_\mu(z^\ast)&\simeq& \Delta\sigma_\tau(z^\ast)
 \nonumber\\&\simeq& -
 \frac{0.528\pi\alpha^2_{\rm em}}{4m^2_W\sin^2{\theta_W}}
 \left( \rho - 1\right)<0.
\end{eqnarray}
Hence, there are three different sign-definite observables to
measure only one parameter of the Abelian $Z'$ boson. This
one-parametric dependence of the Abelian $Z'$ signal gives a
possibility to derive severe model-independent constraints on the
$Z'$ couplings. Thus, the observables $\Delta\sigma_l(z^\ast)$
were found to be the most sensitive for the Abelian $Z'$ searches
when one treats the LEP data.

The present investigation shown that the necessary condition of
the renormalizability of the underlying theory, formulated in
relations (\ref{chiral}) and (\ref{abelian}), gives the
possibility to determine a number of specific correlations between
different scattering processes and to introduce the
model-independent observables convenient in searching for signals
of the chiral and the Abelian $Z'$ bosons both at future $e^+e^-$
colliders and at energies of LEP.

\end{document}